\documentclass[conference]{IEEEtran}
\IEEEoverridecommandlockouts

\usepackage{hyperref}

\usepackage{cite}
\usepackage{amsmath,amssymb,amsfonts}
\usepackage{algorithmic}
\usepackage{graphicx}
\usepackage{tikz}
\usetikzlibrary{positioning}
\usepackage{pgfplots}
\usepackage{textcomp}
\usepackage{xcolor}
\def\BibTeX{{\rm B\kern-.05em{\sc i\kern-.025em b}\kern-.08em
		T\kern-.1667em\lower.7ex\hbox{E}\kern-.125emX}}

\definecolor{TUMblue}{RGB}{0,101,189} %
\definecolor{TUMpantone301}{RGB}{0,82,147}
\definecolor{TUMgreen}{RGB}{0,124,48}
\definecolor{TUMred}{RGB}{196,7,27}
\definecolor{TUMpantone383}{RGB}{162,173,0}
\definecolor{TUMorange}{RGB}{227,114,34}

\definecolor{plot1}{RGB}{0,101,189} %
\definecolor{plot2}{RGB}{196,7,27} %
\definecolor{plot3}{RGB}{227,114,34} %
\definecolor{plot4}{rgb}{0.49400,0.18400,0.55600} %
\definecolor{plot5}{RGB}{162,173,0} %
\definecolor{plot6}{cmyk}{0.65,0.19,0.01,0.04} %

\usepackage{siunitx}
\usepackage[nolist]{acronym}
\usepackage{csquotes}
\usepackage{bm}
\usepackage{mleftright}
\usepackage[only,llbracket,rrbracket]{stmaryrd}

\newcommand{\kfec}{k_{\mathrm{fec}}}
\newcommand{\nfec}{n_{\mathrm{fec}}}
\newcommand{\Rfec}{R_{\mathrm{fec}}}

\newcommand{\kdm}{k_{\mathrm{dm}}}

\newcommand{\Rdm}{R_{\mathrm{dm}}}
\newcommand{\Rsdm}{R_{\mathrm{sdm}}}

\newcommand{\Hmat}{\bm{H}}
\newcommand{\Gmat}{\bm{G}}
\newcommand{\Imat}{\bm{I}}
\newcommand{\Pmat}{\bm{P}}

\newcommand{\uvec}{\bm{u}}
\newcommand{\vvec}{\bm{v}}
\newcommand{\zvec}{\bm{z}}
\newcommand{\pvec}{\bm{p}}
\newcommand{\svec}{\bm{s}}
\newcommand{\pvecbreve}{\bm{{\breve{p}}}}

\newcommand{\Hsys}{\Hmat_{\mathrm{sys}}}

\newcommand{\Hv}{\Hmat_{\text{v}}}
\newcommand{\Hz}{\Hmat_{\text{z}}}
\newcommand{\Hp}{\Hmat_{\text{p}}}

\newcommand{\Htsys}{\transpose{\Hmat}_{\mathrm{sys}}}

\newcommand{\Htv}{\transpose{\Hmat}_{\text{v}}}
\newcommand{\Htz}{\transpose{\Hmat}_{\text{z}}}
\newcommand{\Htp}{\transpose{\Hmat}_{\text{p}}}

\newcommand{\Hpbreve}{\bm{\breve{H}}_{\mathrm{p}}}
\newcommand{\Htpbreve}{\transpose{\bm{\breve{H}}}_{\mathrm{p}}}

\newcommand{\cn}{\mathsf{c}}
\newcommand{\vn}{\mathsf{v}}
\newcommand{\Ncal}{\mathcal{N}}
\newcommand{\neighv}{\mathcal{N}_{\vn}}
\newcommand{\neighc}{\mathcal{N}_{\cn}}

\newcommand{\transpose}[1]{#1^{\mathrm{T}}}

\DeclareMathOperator*{\argmin}{argmin}

\usepackage{mathtools}

\DeclarePairedDelimiter\idxset{\llbracket}{\rrbracket}

\newcommand{\idxsetx}[2]{\idxset{#1\:{:}\:#2}} 
\begin{acronym}
\acro{AWGN}{additive white Gaussian noise}
\acro{AR4JA}{accumulate-repeat-$4$-jagged-accumulate}
\acro{BP}{belief propagation}
\acro{SC}{successive cancellation}
\acro{SCL}{successive cancellation list}
\acro{BMD}{bit-metric decoding}
\acro{SMD}{symbol-metric decoding}
\acro{FEC}{forward error correction}
\acro{PAS}{probabilistic amplitude shaping}
\acro{PMF}{probability mass function}
\acro{PDF}{probability density function}
\acro{ASK}{amplitude shift keying}
\acro{QAM}{quadrature amplitude modulation}
\acro{APSK}{amplitude phase-shift keying}
\acro{LDPC}{low-density parity-check}
\acro{FER}{frame error rate}
\acro{BRGC}{binary reflected Gray code}
\acro{DM}{distribution matcher}
\acro{BICM}{bit-interleaved coded modulation}
\acro{EXIT}{extrinsic information transfer}
\acro{MB}{Maxwell-Boltzmann}
\acro{GMI}{generalized mutual information}
\acro{PAR}{peak-to-average power ratio}
\acro{SE}{spectral efficiency}
\acro{SPB}{sphere packing bound}
\acro{HT}{Hadamard transform}
\acro{SNR}{signal-to-noise ratio}
\acro{DE}{density evolution}
\acro{DifE}{differential evolution}
\acro{MI}{mutual information}
\acro{PSK}{phase-shift keying}
\acro{DOF}{degree of freedom}
\acro{CSI}{channel state information}
\acro{RA}{repeat-accumulate}
\acro{IRA}{irregular repeat-accumulate}
\acro{OP}{operating point}
\acro{PS}{probabilistic shaping}
\acro{GS}{geometric shaping}
\acro{AOP}{achievable operating point}
\acro{NUC}{non-uniform constellation}
\acro{LLR}{log-likelihood ratio}
\acro{iid}{independent and identically distributed}
\acro{MMSE}{minimum-mean square error}
\acro{ZF}{zero forcing}
\acro{MF}{matched filter}
\acro{MIMO}{multiple-input multiple-output}
\acro{BER}{bit error rate}
\acro{SER}{symbol error rate}
\acro{MSE}{mean square error}
\acro{NB-LDPC}{non-binary low-density parity-check}
\acro{BiAWGN}{binary-input additive white Gaussian noise}
\acro{PEG}{progressive edge-growth}
\acro{WDM}{wave-division multiplexing}
\acro{WSS}{wavelength selective switch}
\acro{QPSK}{quadrature phase-shift keying}
\acro{CCDM}{constant composition DM}
\acro{NLIN}{non-linear interference noise}
\acro{EGN}{enhanced Gaussian noise model}
\acro{QS}{quadrant shaping}
\acro{CM}{coded modulation}
\acro{SW}{symbol-wise}
\acro{BW}{bit-wise}
\acro{MUI}{multi-user interference}
\acro{PDM}{product distribution matching}
\acro{DMT}{discrete multitone}
\acro{SVD}{singular value decomposition}
\acro{NB}{non-binary}
\acro{MF}{matched filter}
\acro{ZF}{zero forcing}
\acro{WF}{Wiener filter}
\acro{UE}{user equipment}
\acro{PRBI}{pseudo-random bit interleaver}
\acro{LLR}{log-likelihood ratio}
\acro{RV}{random variable}
\acro{ATSC}{Advanced Television Systems Committee}
\acro{OFDM}{orthogonal frequency division multiplexing}
\acro{DMS}{discrete memoryless source}
\acro{NBBC}{natural based binary code}
\acro{NBC}{natural binary codes}
\acro{OOK}{on-off keying}
\acro{SMDM}{shell mapper as distribution matcher}
\acro{SD-FEC}{soft-decision forward error correction}
\acro{ML}{maximum likelihood}
\acro{EM}{expectation maximization}
\acro{DSP}{digital signal processing}
\acro{SC-LDPC}{spatially coupled low-density parity-check code}
\acro{NGMI}{normalized generalized mutual information}
\acro{GMM}{Gaussian mixture model}
\acro{BEEC}{binary error and erasure channel}
\acro{CN}{check node}
\acro{VN}{variable node}
\acro{MET}{multi edge type}
\acro{TMP}{ternary message passing}
\acro{BMP}{binary message passing}
\acro{QC}{quasi-cylic}
\acro{DSL}{digital subscriber line}
\acro{BPSK}{binary phase shift keying}
\acro{eMBB}{enhanced mobile broadband}
\acro{NA}{normal approximation}
\acro{B-DMC}{binary input discrete memoryless channel}
\acro{TS}{time sharing}
\acro{DD}{direct detection}
\acro{IM}{intensity modulation}
\acro{PAM}{pulse amplitude modulation}
\acro{bpcu}{bit per channel use}
\acro{B2B}{back-to-back}
\acro{SSMF}{standard single-mode fiber}
\acro{CD}{chromatic dispersion}
\acro{BCJR}{Bahl-Cocke-Jelinek-Raviv}
\acro{AWG}{arbitrary waveform generator}
\acro{MZM}{Mach-Zehnder modulator}
\acro{VOA}{variable optical attenuator}
\acro{PD}{photodetector}
\acro{DSO}{digital storage oscilloscope}
\acro{AIR}{achievable information rate}
\acro{DA}{driver amplifier}
\acro{ISI}{inter-symbol interference}
\acro{i.i.d.}{independent and identically distributed}
\acro{MLHY}{multi-level Honda-Yamamoto}
\acro{LLPS}{linear layered probabilistic shaping}
\acro{SDM}{syndrome distribution matcher}
\acro{MAP}{maximum a posteriori}
\acro{APP}{a posteriori probability}
\acro{PPS}{probabilistic parity shaping}
\acro{MCSDM}{minimum cost SDM}
\acro{G-SDM}{$\boldsymbol{G}$-based SDM}
\acro{SE-SDM}{sequential encoding SDM}
\acro{SBE-SDM}{sequential block encoding SDM}
\acro{RCU}{random coding union}
\end{acronym}

\begin{document}

\title{Efficient Probabilistic Parity Shaping for Irregular Repeat-Accumulate LDPC Codes
\thanks{This work was supported by the National Science Foundation (NSF) grant 1911166. Any opinions, findings, and conclusions or recommendations expressed in this material are those of the authors and do not necessarily reflect the views of the NSF.}
}

\author{\IEEEauthorblockN{Diego Lentner}%
\IEEEauthorblockA{
\textit{Technical University of Munich}\\
Munich, Germany \\
diego.lentner@tum.de}
\and
\IEEEauthorblockN{Thomas Wiegart}%
\IEEEauthorblockA{
\textit{Technical University of Munich}\\
Munich, Germany \\
thomas.wiegart@tum.de}
\and
\IEEEauthorblockN{Richard D. Wesel}%
\IEEEauthorblockA{
\textit{University of California Los Angeles}\\
Los Angeles, USA \\
wesel@ucla.edu}
}

\maketitle

\begin{abstract}
Algorithms are presented that efficiently shape the parity bits of systematic irregular repeat-accumulate (IRA) low-density parity-check (LDPC) codes by following the sequential encoding order of the accumulator. Simulations over additive white Gaussian noise (AWGN) channels with on-off keying show a gain of up to 0.9 dB over uniform signaling.
\end{abstract}


\section{Introduction}
\label{sec:intro}

\Ac{PS} improves the power efficiency of communication by modifying the distribution of symbols; see \cite[Sec.~II]{bocherer2015bandwidth} for a list of \ac{PS} schemes. We focus on \ac{PAS} \cite{bocherer2015bandwidth}, \cite{bocherer2023probabilistic}, which is a layered scheme with an outer \ac{DM} and an inner systematic \ac{FEC} encoder. For example, for long codes one might use a \ac{CCDM} \cite{schulte2016constant} and a systematic \ac{LDPC} encoder \cite{gallager1960low}, \cite{mackay1995good}. The \ac{DM} maps data bits to a variable $A$ with probability mass function (pmf) $P_A$, typically representing the distribution of amplitudes, and outputs bits fed to a systematic FEC encoder. The encoder outputs parity bits with an (almost) uniform pmf $P_S$, which typically control the sign $S$ of a channel input symbol $X=S\cdot A$. \Ac{PAS} can approach channel capacity if the optimal input distribution factors as $P_X = P_S \cdot P_A$. Two key features of \Ac{PAS} are that it enables flexible rate adaptation by adjusting the \ac{DM} pmf, and its layered architecture can be used with any \ac{DM} and systematic \ac{FEC} encoder.

\Ac{LLPS} extends \ac{PAS} by also shaping the parity bits, which is called \ac{PPS} \cite{bocherer2019llps}. This enables communication with any $P_X$. \Ac{PPS} uses a nonlinear \ac{SDM} that encodes into cosets with the desired distribution. 
\ac{PPS} algorithms for \ac{LDPC} codes can be based on \ac{BP}-like \ac{SDM} algorithms and the generator matrix $\Gmat$ \cite{wiegart2023probabilistic}. This guarantees that the output is a valid codeword, which improves previous \ac{BP}-based approaches such as \cite{mondelli2018how}. However, complexity is a limiting factor for long codes.

Our contributions are as follows. We first reformulate the \ac{LLPS} framework to facilitate \ac{PPS} design. Second, we present a sequential \ac{SDM} algorithm for \ac{IRA} codes. The algorithm has encoding complexity $\mathcal{O}(\nfec)$ and efficiently shapes all bits for long codes such as the DVB-S2 \ac{LDPC} codes \cite{dvbs2} of length $\num{64800}$, which was not previously possible. We present a block encoding variant of the \ac{SDM} that improves performance but increases complexity. Third, we propose a simple strategy to select the shaping bits. We evaluate the performance over \ac{AWGN} channels with \ac{OOK} and compare to the \ac{SDM} in \cite{wiegart2023probabilistic}, the \ac{TS} scheme in \cite{git2019protographbased}, and polar-coded \ac{PS} in \cite{honda2013polar,wiegart2019shaped}.

This paper is organized as follows.
Sec.~\ref{sec:prelim} reviews \ac{LLPS}, the $\boldsymbol{G}$-based SDM from \cite{wiegart2023probabilistic}, and \ac{IRA} codes. Sec.~\ref{sec:llps-revisited} reformulates \ac{LLPS} to provide a new perspective. Sec.~\ref{sec:main} presents our sequential encoding algorithm, its block encoding variant, the shaping bit selection strategy, and a complexity analysis. Sec.~\ref{sec:simulations} presents numerical results. Sec.~\ref{sec:conclusion} concludes the paper.

\section{Preliminaries}
\label{sec:prelim}

Vectors and matrices are written with lowercase and uppercase bold symbols $\bm{x}$ and $\bm{X}$, respectively. We use the index set notation $\idxsetx{i}{j}\triangleq \lbrace i,\ldots,j \rbrace$, $i\leq j$, and $\idxset{n}\triangleq\idxsetx{1}{n}$. We write $\transpose{\bm{a}}_j$ for the $j$-th column of $\bm{A}$, and $[\bm{A}]_{\mathcal{I},\mathcal{J}}$ to denote the submatrix of $\bm{A}$ with row and column indices in $\mathcal{I}$ and $\mathcal{J}$, respectively. $\Imat_r$ denotes the $r\times r$ identity matrix.

\subsection{Linear Layered Probabilistic Shaping (LLPS)}
\label{subsec:LLPS}
\Ac{LLPS} \cite{bocherer2019llps} maps data bits to codewords of a linear code, where the codewords have a non-uniform distribution. Consider a linear code of length $\nfec$ and dimension $\kfec$, and let $m=\nfec-\kfec$. Suppose a $m\times\nfec$ parity-check matrix is
\begin{align}
    \Hmat = \begin{bmatrix} \Hsys & \Hp \end{bmatrix} 
    \label{eq:H_decomp_Hsys_Hp}
\end{align}
where $\Hsys$ is $m\times\kfec$, and  $\Hp$ is full-rank and square. For $\ell \in \idxsetx{0}{\kfec}$, the paper \cite{bocherer2019llps} partitions $\Hmat$ into a $m\times(\kfec-\ell)$ matrix $\Hv$ and a $m\times(m+\ell)$ matrix $\Hpbreve$ so that 
\begin{align}
    \Hmat = \begin{bmatrix} \Hv & \Hpbreve \end{bmatrix} \,. 
    \label{eq:H_decomp_llps}
\end{align}
As for \ac{PAS}, the \ac{DM} output is systematically encoded. Let $\vvec$ be the \ac{DM} output with $\kfec-\ell$ bits. \ac{FEC} encoding is performed in two steps: compute the syndrome $\svec_{\mathrm{v}} = \vvec\Htv$ and then compute a solution $\pvecbreve$ of $\pvecbreve\Htpbreve=\svec_{\mathrm{v}}$. The encoder output is $[\vvec|\pvecbreve]$, which is a valid codeword since $[\vvec|\pvecbreve]\transpose{\Hmat}=\mathbf{0}$. Choosing $\ell=0$ recovers \ac{PAS}.

The paper \cite{bocherer2019llps} describes a \ac{MCSDM} that computes
\begin{align}
    \pvecbreve = \argmin_{\bm{\tilde{p}}\in\lbrace0,1\rbrace^{m+\ell}} f(\bm{\tilde{p}}) \quad\text{s.t.}\quad \bm{\tilde{p}}\Htpbreve = \svec_{\mathrm{v}}
    \label{eq:LLPS_MCSDM}
\end{align}
where $f(.)$ is a cost function, e.g., the Hamming weight. 
The search space in \eqref{eq:LLPS_MCSDM} is reduced to $\lbrace0,1\rbrace^{\ell}$ by transforming $\Hpbreve$.
However, its exponential complexity in $\ell$ makes the \ac{MCSDM} infeasible for most applications.

\subsection{\texorpdfstring{$\boldsymbol{G}$}{G}-based SDM for LDPC Codes}
\label{subsec:G-shaping}
\Ac{LDPC} codes are linear block codes with a sparse parity-check matrix $\Hmat$ commonly represented by a Tanner graph, where \acp{VN} and \acp{CN} represent the code symbols and parity check constraints, respectively. Decoding is usually performed via \ac{BP} on this Tanner graph.

The paper \cite{wiegart2023probabilistic} proposed a greedy \ac{SDM} using \ac{BP} on the Tanner graph of a systematic generator matrix $\Gmat=[\Imat_{\kfec}|\Pmat]$. 
We refer to this method as \ac{G-SDM} and distinguish three types of \acp{VN}: message, shaping, and parity \acp{VN}, associated with the \ac{DM} output $\vvec$ of length $(\kfec-\ell)$, the $\ell$ shaping bits $\zvec$, and the $m$ parity bits $\pvec$, respectively.
After each \ac{BP} iteration, the most reliable shaping \ac{VN} is fixed (decimation step) so the algorithm terminates after $\ell$ iterations.
Recall that the shaping \acp{VN} represent systematic code bits. By limiting the decimation choice to such \acp{VN}, the algorithm guarantees encoding to a valid codeword.

\textit{Remark.} The Tanner graph of $\Gmat$ is equivalent to the Tanner graph of the parity-check matrix $\Hmat = [\transpose{\Pmat}|\Imat_m]$. 
To apply the \ac{G-SDM} to general \ac{LDPC} codes with $\Hmat$ as in \eqref{eq:H_decomp_Hsys_Hp}, we compute
\begin{align}
    \Hmat' = [\Hsys\Hp^{-1}|\Imat_m]\,.
\end{align}
However, $\Hsys\Hp^{-1}$ is not necessarily sparse. The density of $\Hp^{-1}$ typically increases as $m$ grows. The paper \cite{wiegart2023probabilistic} notes that $\Gmat$ and $\Hsys\Hp^{-1}$ should not be too dense.

\subsection{Systematic IRA LDPC Codes}
\label{subsec:IRA}
Systematic \ac{RA} \cite{divsalar1998coding} and \ac{IRA} \cite{jin2000irregular}, \cite{yang2004design} codes are \ac{LDPC} codes with $\Hmat= [\Hsys | \Hp]$ where $\Hp$ is the $m\times m$ square dual-diagonal matrix
\begin{align}
    \Hp = \begin{bmatrix}
        1 & & & \\
        1 & 1& & \\
        & \ddots & \ddots & \\
        & & 1 & 1
    \end{bmatrix} \,.
    \label{eq:Hp_IRA}
\end{align}
This structure enables efficient accumulator-based encoding. Let $\svec=\uvec\Htsys$ and $p_1=s_1$. For $t\in\idxsetx{2}{m}$, the parity bits $p_t$ are sequentially computed as
\begin{align}
    p_t = s_t + p_{t-1} \,.
    \label{eq:IRA_accumulator_pt}
\end{align}
We write $\pvec = \mathsf{Acc}(\svec,0)$ to denote the accumulator output with input $\svec$ and initial state 0. The final codeword is $[\uvec|\pvec]$. Clearly, $\svec=\pvec \Htp$, and $[\uvec|\pvec]\transpose{\Hmat}=\bm{0}$, i.e., $[\uvec|\pvec]$ is a codeword.

\section{LLPS Revisited}
\label{sec:llps-revisited}
Partitioning $\Hmat$ as in \eqref{eq:H_decomp_llps} helps to derive achievable rates for \ac{LLPS} \cite{bocherer2019llps}. However, focusing on $\Hpbreve$ can be misleading when designing efficient \acp{SDM}. For example, encoding into cosets of $\Hpbreve$ for \ac{MCSDM} requires storing $2^\ell$ vectors of length $2^{\ell+m}$.

The paper \cite{wiegart2023probabilistic} instead considered the systematic $\Gmat$ and points out that reserving $\ell$ systematic bits for shaping the parity bits is equivalent to a \ac{SDM}. However, focusing on $\Gmat$ diverts attention from the key insight that encoding into cosets of $\Hpbreve$ is efficient if there is an efficient encoder for $\Hmat$.\footnote{The problem of efficiently selecting \enquote{good} shaping bits remains.}

We bridge the two perspectives. Suppose $\Hmat$ is systematic and partition $\Hsys$ to give
\begin{align}
    \Hmat = \Big[\begin{matrix} \mathrlap{\overbrace{\phantom{\begin{matrix}\Hv & \Hz\end{matrix}}}^{\Hsys}}\Hv & \mathrlap{\underbrace{\phantom{\begin{matrix}\Hz & \Hp\end{matrix}}}_{\Hpbreve}}\Hz & \Hp \end{matrix}\Big]
    \label{eq:H_decomp_llps_revisited}
\end{align}
with the $m\times(\kfec-\ell)$ matrix $\Hv$, $m\times\ell$ matrix $\Hz$, and full-rank $m\times m$ matrix $\Hp$. 
For an input $[\vvec|\zvec]$ with $\kfec$ bits, a systematic encoder computes the unique solution $\pvec$ of
\begin{align}
    [\vvec|\zvec|\pvec]\transpose{\Hmat} &= \svec_{\mathrm{vz}}+\pvec\Htp = \mathbf{0}.
\end{align}
where $\svec_{\mathrm{vz}} = \vvec\Htv+\zvec\Htz$. We write $\pvec=\mathsf{\Phi}(\svec_{\mathrm{vz}})$ to emphasize that $\pvec$ is a function of $\svec_{\mathrm{vz}}$. For example, an efficient encoder for \ac{IRA} codes is $\mathsf\Phi(\svec_{\mathrm{vz}}) = \mathsf{Acc}(\svec_{\mathrm{vz}},0)$ as in \eqref{eq:IRA_accumulator_pt}.

Given $\svec_{\mathrm{v}} = \vvec\Htv$, the task of the \ac{SDM} is determining a $\zvec$ so that $[\zvec|\mathsf{\Phi}(\svec_{\mathrm{v}}+\zvec\Htz)]$ has desired properties. For example, a \ac{MCSDM} computes
\begin{align}
    \zvec &= \argmin_{\bm{\tilde{z}}\in\lbrace0,1\rbrace^{\ell}} f\mleft(\left[\bm{\tilde{z}}|\mathsf{\Phi}(\svec_{\mathrm{v}}+\bm{\tilde{z}}\Htz)\right]\mright)
    \label{eq:LLPS_revisited_MCSDM}
\end{align}
for a cost function $f(.)$ and returns $[\zvec|\mathsf{\Phi}(\svec_{\mathrm{v}}+\zvec\Htz)]$. The reduced search space $\lbrace0,1\rbrace^{\ell}$ is a consequence of \eqref{eq:H_decomp_llps_revisited},
and the \ac{MCSDM} benefits from an efficient parity encoder $\mathsf{\Phi}(.)$. 

We summarize \ac{LLPS}. The \ac{DM} input is a vector of dimension $\kdm$ and its output is $\vvec$. The \ac{SDM} computes $\svec_{\mathrm{v}} = \vvec\Htv$ and outputs a shaping string $\zvec$ and the correponding parity bits $\pvec=\mathsf{\Phi}(\svec_{\mathrm{v}}+\zvec\Htz)$. The codeword is $[\vvec|\zvec|\pvec]$ and satisfies $[\vvec|\zvec|\pvec]\transpose{\Hmat}=\mathbf{0}$.
The overall transmission rate is
\begin{align}
    R_{\mathrm{LLPS}} = \frac{\kdm}{\nfec}
    &= \Rdm - \frac{\Rdm}{\Rsdm}(1-\Rfec) \label{eq:rate_llps}
\end{align}
where $\Rdm = \frac{\kdm}{\kfec-\ell}$,  $\Rsdm=\frac{m}{m+\ell}$, and $\Rfec=\frac{\kfec}{\nfec}$.

\section{Sequential PPS for IRA LDPC Codes}
\label{sec:main}

Consider a $m\times \nfec$ parity-check matrix $\Hmat$ as in \eqref{eq:H_decomp_llps_revisited}. We associate CN $i\in\idxset{m}$ with the $i$-th row of $\Hmat$, and VN $j\in\idxset{\nfec}$ with its $j$-th column. Let $\mathcal{V}$, $\mathcal{Z}$ and $\mathcal{P}$ be the index sets corresponding to the columns of $\Hmat$ that form the submatrices $\Hv$, $\Hz$ and $\Hp$, respectively. The index sets partition $\idxset{\nfec}$. Let $\neighc(i)$ and $\neighv(j)$ be the neighborhoods of \ac{CN} $i$ and \ac{VN} $j$, respectively.
We call \ac{CN} $i$ the \textit{pivot \ac{CN}} of \ac{VN} $j$ if CN $i$ is the first CN that VN $j$ is connected to, i.e., $i\in\neighv(j)$ and $\idxset{i-1} \cap \neighv(j) = \emptyset$.
Let $\mathcal{A}\subseteq\idxset{\nfec}$ be any subset of column indices. Let $\mathcal{A}_i$ be the set of all \acp{VN} in $\mathcal{A}$ that share the same pivot \ac{CN} $i$, and let $\mathcal{A}_i=\emptyset$ if CN $i$ is not the pivot CN of any VN in $\mathcal{A}$. The sets $\mathcal{A}_i$ partition $\mathcal{A}$. The index set of all pivot \acp{CN} w.r.t. $\mathcal{A}$ is $\Gamma(\mathcal{A})$, and $|\Gamma(\mathcal{A})|\leq |\mathcal{A}|$. 

Consider $\Hp$ as in \eqref{eq:Hp_IRA}, i.e., $\Hmat$ describes a systematic \ac{IRA} code. Let $i_{t}$, $i_{t+1}$ be two subsequent pivot \acp{CN} in $\Gamma(\mathcal{Z})$ w.r.t. the shaping \acp{VN} $\mathcal{Z}$. We define the $t$-th encoding section as
\begin{align}
    \mathcal{B}_t= \begin{cases}
        \idxsetx{1}{i_{t+1}-1} & t = 1 \\
        \idxsetx{i_{t}}{i_{t+1}-1} & t \in \idxsetx{2}{|\Gamma(\mathcal{Z})|} \\
        \idxsetx{i_{t}}{m} & t = |\Gamma(\mathcal{Z})| 
    \end{cases}
\end{align}
which corresponds to the intermediate \acp{CN} between the pivot \acp{CN} $i_{t}$, $i_{t+1}$, and to the parity \acp{VN} that connect them through the characteristic \ac{IRA} zigzag chain.

\subsection{Sequential Encoding SDM}
\label{subsec:SE-SDM}
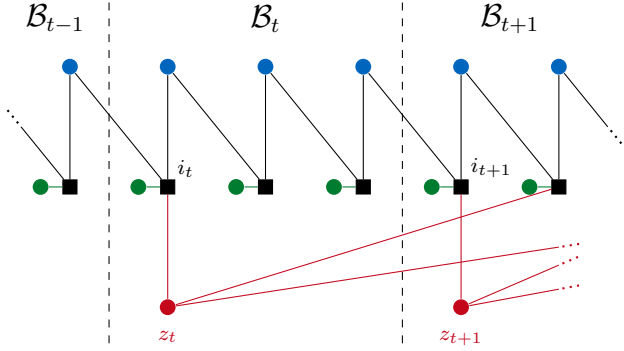
\begin{figure}
    \centering
    \begin{tikzpicture}
\usetikzlibrary{arrows}  

\tikzset{%
          insert new path/.style={%
             insert path={%
                  node[pos=0.7,sloped]{\tikz \draw[#1,thick] (-.2pt,0) -- ++(.2 pt,0);}
                  }
             }
         }
\begin{scope}[yscale=0.8,xscale=0.65, a/.style = {insert new path = {-triangle 90}}] \footnotesize

	\foreach \i [evaluate=\i as \ii using \i*2] in {0,...,5} {
		\node[draw, shape=rectangle, fill, minimum width=0.25, minimum height=0.25, anchor=center] (cn_\i) at ([shift={(0.5,0)}]\ii, 2) {};
        \node[fill, TUMgreen, circle, inner sep=0, minimum size=0.21cm] (datavn_\i) at ([shift={(-0.6,0)}]cn_\i) {};
		\draw[-, TUMgreen] (datavn_\i) -- (cn_\i);
	}
	
	\foreach \i [evaluate=\i as \ii using \i*2] in {0,...,5} {
            \node[fill, TUMblue, circle, inner sep=0, minimum size=0.21cm] (parityvn_\i) at ([shift={(0.5,0)}]\ii,4) {};
			\draw[-] (cn_\i) -- (parityvn_\i);	
	}

    \node[above right=-0.1cm of cn_1] {$i_t$};
    \node[above right=-0.1cm of cn_4] {$i_{t+1}$};

    \draw[-] (parityvn_0) -- (cn_1);
	\draw[-] (parityvn_1) -- (cn_2);
	\draw[-] (parityvn_2) -- (cn_3);
	\draw[-] (parityvn_3) -- (cn_4);
	\draw[-] (parityvn_4) -- (cn_5);
    \draw[-] (parityvn_5) --++ (1,-1) node[inner sep=0.04cm] (contline1) {};
    \draw[thick, dotted] (contline1) --++ (0.3,-0.3);
    \draw[-] (cn_0) --++ (-1, 1) node[inner sep=0.04cm] (contline2) {};
    \draw[thick, dotted] (contline2) --++ (-0.3,0.3);
	
    \node[fill,TUMred,circle,inner sep=0,minimum size=0.21cm] (svn_1) at (2.5, 0) {};
    \node[fill,TUMred,circle,inner sep=0,minimum size=0.21cm] (svn_2) at (8.5, 0) {};
    \node[below=0.1cm of svn_1, TUMred] {$z_{t}$};
    \node[below=0.1cm of svn_2, TUMred] {$z_{t+1}$};
	
	\draw[TUMred] (svn_1) -- (cn_1);
	\draw[TUMred] (svn_1) -- (cn_5);
    \draw[TUMred] (svn_1) --++ (8, 1) node[inner sep=0.04cm] (contline3) {};
    \draw[TUMred, thick, dotted] (contline3) --++ (0.5,0.0625);
	
	\draw[TUMred] (svn_2) -- (cn_4);
    \draw[TUMred] (svn_2) --++ (2, 0.3)  node[inner sep=0.04cm] (contline4) {};
    \draw[TUMred, thick, dotted] (contline4) --++ (0.5,0.075);
    \draw[TUMred] (svn_2) --++ (2, 0.7)  node[inner sep=0.04cm] (contline5) {};
    \draw[TUMred, thick, dotted] (contline5) --++ (0.4,0.14);

    \draw[dashed] (1.3,-0.6) -- (1.3, 5.2);
    \draw[dashed] (7.3,-0.6) -- (7.3, 5.2);

    \node at (4.5, 4.8) {\large$\mathcal{B}_t$};
    \node at (0.2, 4.8) {\large$\mathcal{B}_{t-1}$};
    \node at (9.5, 4.8) {\large$\mathcal{B}_{t+1}$};

\end{scope}
\end{tikzpicture}
    \caption{Simplified SDM Tanner graph of an IRA code.}
    \label{fig:tanner_ira}
\end{figure}

Suppose all shaping \acp{VN} in $\mathcal{Z}$ have distinct pivot \acp{CN}, i.e., $|\Gamma(\mathcal{Z})|=\ell$ and $|\mathcal{Z}_t|=1$ for all $t\in\idxset{\ell}$.
Without loss of generality, order the columns $\transpose{\bm{h}}_{\mathrm{z},t}$ of $\Hz$ in ascending pivot \ac{CN} order.
We propose a \ac{SE-SDM} inspired by the accumulator in Sec.~\ref {subsec:IRA}. Let the input to the \ac{SDM} be $\svec_{\mathrm{v}} = \vvec\Htv$, where $\vvec$ is the shaped output of a \ac{DM}.

Initialize $\svec\leftarrow\svec_{\mathrm{v}}$. For each pivot \ac{CN} $i_t \in \Gamma(\mathcal{Z})$, $t\in\idxset{\ell}$:
\begin{enumerate}
    \item Select the $t$-th shaping bit $z_t$ such that
    \begin{align}
        z_t = \argmin_{\tilde{z}\in\lbrace0,1\rbrace} f\mleft(\mleft[\tilde{z}\middle|\bm{\psi}(\tilde{z})\mright] \mright) 
        \label{eq:SE-SDM}
    \end{align}    
    where 
    \begin{align*}
        \bm{\psi}(\tilde{z}) = 
        \begin{cases} 
        \mathsf{Acc}\mleft(\mleft[\svec\mright]_{\mathcal{B}_t} + \tilde{z} [\bm{h}_{\mathrm{z},t}]_{\mathcal{B}_t}, 0\mright) & t=1\\
        \mathsf{Acc}\mleft(\mleft[\svec\mright]_{\mathcal{B}_t} + \tilde{z} [\bm{h}_{\mathrm{z},t}]_{\mathcal{B}_t}, p_{i_t-1}\mright) & t\in\idxsetx{2}{\ell}
        \end{cases}
    \end{align*} 
    If there is more than one minimizer $\tilde{z}$, set $z_t=0$.
    \item Update $\svec\leftarrow\svec+z_t \bm{h}_{\mathrm{z},t}$ and $[\pvec]_{\mathcal{B}_t} \leftarrow \bm{\psi}(z_t)$.
\end{enumerate}
Finally, output $[\bm{z}|\bm{p}]$. The parity bits satisfy $\bm{p}=\mathsf{\Phi}(\svec_{\mathrm{v}}+\zvec\Htz)$ and we have $[\vvec|\zvec|\pvec]\transpose{\Hmat}=\mathbf{0}$.

The algorithm can support shaping \acp{VN} that share a pivot \ac{CN} $i_t$. These \acp{VN} share the encoding section $\mathcal{B}_t$ and must be chosen jointly to avoid overwriting previous decisions. However, the search space of \eqref{eq:SE-SDM} increases exponentially in $|\mathcal{V}_{i_t}|$, and numerical experiments did not show improvements. We thus design the shaping set to have unique pivot \acp{CN} via the selection strategy in Sec.~\ref{subsec:shaping_bit_selection}.

Fig.~\ref{fig:tanner_ira} depicts a section of the simplified \ac{SDM} Tanner graph for an \ac{IRA} code, where the subgraph corresponding to $\Hv$ is replaced by degree-1 syndrome \acp{VN} (green) associated with $\svec_{\mathrm{v}}=\vvec\Htv$. The accumulator \eqref{eq:IRA_accumulator_pt} runs from left to right along the zigzag chain and successively outputs the parity bits (blue). Consider the $t$-th pivot \ac{CN} $i_t$ and its encoding section $\mathcal{B}_t$. Given $p_{i_{t}-1}$, we compute the parity bits $[\pvec]_{\mathcal{B}_t}$ for both $z_t=0$ and $z_t=1$ and choose the lowest cost output. After $z_t$ is fixed, we remove the shaping \ac{VN} (red) from the graph and update the syndrome. We continue with the next pivot \ac{CN}.

\textit{Remark.} The tiebreaker in Step 1) outperforms a randomized rule and is motivated by the following observation. 
Systematically encoded parity bits are approximately uniform, even when the input is shaped \cite[Sec.~IV-A2]{bocherer2015bandwidth}. However,
the syndromes $\svec_{\mathrm{v}}=\vvec\Htv$ are typically sparse if $\vvec$ was shaped by a \ac{DM}. An accumulator fed with sparse input will produce large runs of zeros and ones, which can be efficiently flipped by a few shaping bits. Our tie-breaking rule aims to keep the syndrome sparse by adding only a few columns $\bm{h}_{\mathrm{z},t}$. 

\textit{Remark.} The \ac{FEC} decoder does not change for \ac{SDM}: the decoder computes the estimate $[\bm{\hat{v}}|\bm{\hat{z}}|\bm{\hat{p}}]$ and discards $\bm{\hat{z}}$ and $\bm{\hat{p}}$. It only needs to know the positions $\mathcal{Z}$. 

\subsection{Shaping Bit Selection}
\label{subsec:shaping_bit_selection}

\ac{LLPS} does not specify the shaping bit positions. Optimizing over all $\binom{\kfec}{\ell}$ choices is generally infeasible. To address this, consider how the \ac{SE-SDM} typically works: in each encoding section, one shaping \ac{VN} effectively either keeps ($z_t{=}0$) or flips ($z_t{=}1$) the accumulator output. Small encoding sections can better approximate the desired output distribution because long encoding sections yield almost uniformly distributed parity bits. We thus propose a selection strategy that makes encoding sections almost equally large. 

Initially, we have $\mathcal{V}=\idxset{\kfec}$ and $\mathcal{Z}=\emptyset$.
Let $i_{\max} =\max \Gamma(\mathcal{V})$ be the last pivot \ac{CN} w.r.t. $\Hsys$. Partition the index set $\idxset{i_{\max}}$ into $\ell$ bins of equal width. Let the first pivot \ac{CN} in a bin have index $i$. Among all \acp{VN} in $\mathcal{V}_i$, randomly remove one \ac{VN} from $\mathcal{V}$ and add it to $\mathcal{Z}$. If there was no pivot \ac{CN} in one bin, decrease $\ell$ by one and continue with the next bin. The algorithm terminates when all bins have been processed and outputs the final partitions $\mathcal{V}$ and $\mathcal{Z}$, which define the matrices $\Hv$ and $\Hz$, respectively. Note that the final value of $\ell$ might deviate from the target number of shaping \acp{VN}.

\subsection{Sequential Block Encoding SDM} 
One can improve the sequential \ac{SDM} by making joint decisions on blocks of $\tau$ shaping \acp{VN}; we call this a \ac{SBE-SDM}. The encoding sections $\mathcal{B}_t$ are merged to contain $\tau$ pivot \acp{CN}. The search space of \eqref{eq:SE-SDM} increases to all $\bm{\tilde{z}} \in \lbrace 0,1 \rbrace^\tau$, and we process $\tau$ columns of $\Hz$ at the same time. The tiebreak rule is generalized to choose the $\bm{\tilde{z}}$ of lowest Hamming weight.

\textit{Remark.} The proposed block encoding can be interpreted as individually performing \ac{MCSDM} for each block encoding section. \ac{SBE-SDM} implements \ac{MCSDM} when $\tau=\ell$.

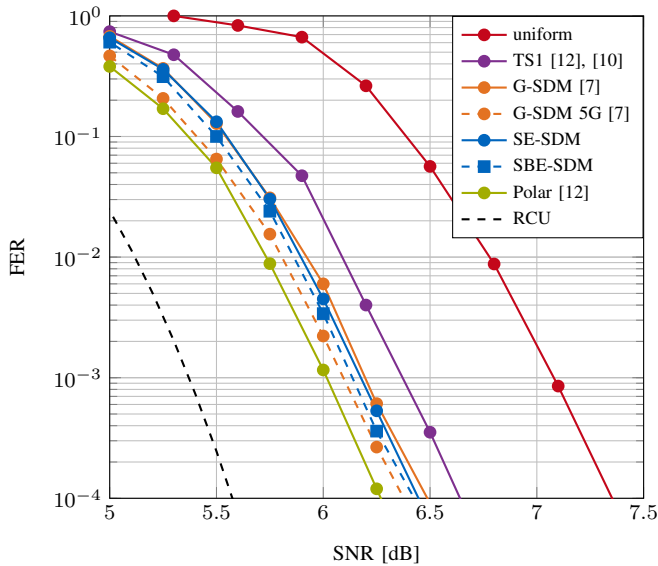
\begin{figure}[t]
    \centering
    \begin{tikzpicture} %
	\begin{axis}[
		height=8cm,
		width=.98\linewidth,
		grid=both,
		xlabel={SNR [dB]},
		ylabel={FER},
		ymode=log,
		xmin=5,
		xmax=7.5,
		ymin=1e-4,
		ymax=1,
		legend cell align={left},
		legend style={at={(1,1)},anchor=north east, font=\scriptsize} %
	]

    \footnotesize

    \addplot[thick, plot2, mark=*] table [x=SNR, y=FER] {res/n1056/wimax_uni.txt};
    \label{plot:r067:wimax_uniform}
    \addlegendentry{uniform} 
    
    \addplot[thick, plot4, mark=*] table [x=SNR, y=FER] {res/n1056/wimax_ts1.txt};
    \label{plot:r067:wimax_TS}
    \addlegendentry{TS1 \cite{wiegart2019shaped}, \cite{git2019protographbased}}

    \addplot[thick, plot3, mark=*] table [x=SNR, y=FER] {res/n1056/wiegartDM_wimax_n1056_34B_s10.txt};
    \label{plot:r067:G-SDM_Wimax}
    \addlegendentry{\ac{G-SDM}\cite{wiegart2023probabilistic}}

    \addplot[thick, plot3, dashed, mark=*, mark options={solid}] table [x=SNR, y=FER] {res/n1056/wiegartDM_5G_n1008_0.78_s32.txt};
    \label{plot:r067:G-SDM_5G}
    \addlegendentry{\ac{G-SDM} 5G\cite{wiegart2023probabilistic}}

    \addplot[thick, plot1, mark=*] table [x=SNR, y=FER] {res/n1056/peg3_nShpBits=10_nJointShp=1_C2_pp_bu_4b.txt};
    \label{plot:r067:SE-SDM}
    \addlegendentry{SE-SDM}

    \addplot[thick, plot1, dashed, mark=square*, mark options={solid}] table [x=SNR, y=FER] {res/n1056/peg3_nShpBits=10_nJointShp=5_C2_pp_bu_4b.txt};
    \label{plot:r067:SBE-SDM}
    \addlegendentry{SBE-SDM}

    \addplot[thick, plot5, mark=*] table [x=SNR, y=FER] {res/n1056/polar_res_n=1024_k=686_dSNR=6.00dB_nShp=110_pXtarget=0.72_Lenc=32_Ldec=32_CRC=16_pest_samples=1000000.txt};
    \label{plot:r067:polar}
    \addlegendentry{Polar \cite{wiegart2019shaped}}
    
    \addplot[thick, black, dashed] table [x=SNR, y=FER] {res/n1056/rcu_shp.txt};
    \label{plot:r067:RCU}
    \addlegendentry{RCU}
		
	\end{axis}
\end{tikzpicture}%
    \caption{\acp{FER} with overall rate $R=2/3$ and $\nfec\approx 1000$.}
    \label{fig:fer_R=0.67}
\end{figure}

\subsection{Complexity Analysis}

\ac{SE-SDM} has linear-time encoding complexity $\mathcal{O}(\nfec)$ for any number $\ell$ of shaping bits, similar to \ac{IRA} encoding \cite{jin2000irregular}. More precisely, \ac{SE-SDM} is approximately twice as complex as encoding the underlying \ac{IRA} code: consider Sec.~\ref{subsec:IRA} and observe that \eqref{eq:SE-SDM} considers disjoint encoding sections and can be implemented by two parallel accumulators to encode section $\mathcal{B}_t$ at time $t$. The overall complexity to compute $\svec$ remains unchanged, and a little overhead is needed to evaluate and compare the objective $f$, e.g., the Hamming weight, within one encoding section.

\Ac{SBE-SDM} applies $2^{\tau}$ parallel accumulators over the $\ell/\tau$ disjoint encoding sections, where again at time $t$ the accumulators run only over encoding section $\mathcal{B}_t$. The overall encoding complexity is thus $\mathcal{O}(2^{\tau}\nfec)$ for any $\ell$.

\ac{G-SDM} encoding \cite{wiegart2023probabilistic} instead requires $\ell$ \ac{BP} iterations with complexity $\mathcal{O}(\ell\nfec)$, and $\ell$ typically grows linearly with $m$ \cite[Eq.~(6)]{bocherer2019llps}. For fixed $\Rfec$, $\Rdm$ and $\Rsdm$, the encoding complexity of \ac{G-SDM} thus scales as $\mathcal{O}(\nfec^2)$, which can be impractical for long codes even if the sparsity constraints from Sec.~\ref{subsec:G-shaping} are met. Note also that \ac{SE-SDM} and \ac{SBE-SDM} have a fixed encoding schedule that simplifies hardware implementation. Decoder-based schemes such as \ac{G-SDM} instead process the shaping bit positions in different order for each input sequence.

\subsection{Generalizations}
\ac{SE-SDM} and \ac{SBE-SDM} apply to unstructured or structured \ac{IRA} codes if $\Hp$ has the form \eqref{eq:Hp_IRA}; extending to generalized \ac{IRA} codes \cite{liva2005simple}, \cite{johnson2005constructions} is straightforward. \ac{SE-SDM} and \ac{SBE-SDM} can be further generalized to \ac{LDPC} codes with (block) lower-triangular $\Hp$, permitting sequential encoding via forward substitution. 
It is unclear how to extend to codes where $\Hp$ is not (block) lower triangular, as connections to previous encoding sections will flip all subsequent, already decided parity bits. This also applies to \ac{IRA}-like codes such as the WiMAX codes \cite{ieee_wimax}. While they can be encoded efficiently using an accumulator, all systematic bits, including the shaping bits, must be fixed initially. This problem is also encountered for the encoding in \cite{richardson2001efficient}, and a resolution would open \ac{SE-SDM} and \ac{SBE-SDM} to a broad class of \ac{LDPC} codes.

\section{Numerical Results}
\label{sec:simulations}

Consider transmission over the \ac{AWGN} channel $Y=X+N$, $N\sim\Ncal(0,\sigma^2)$, with \ac{OOK} modulation, i.e., $A>0$ and $X$ is either $0$ or $A$. The \ac{SNR} is
\begin{align}
    \mathsf{SNR} = \frac{P_X(A)\cdot A^2}{\sigma^2} \,.
\end{align}
Under the constraint $\mathbb{E}[X^2]\leq P$, choosing $P_X(0)>1/2$ increases energy efficiency; see~\cite{wiegart2019shaped}. 
We apply the labeling $b : \lbrace 0,A \rbrace \to \lbrace 0,1 \rbrace$ with $b(0) = 0$ and $b(A)=1$ to map $X$ to code bits and vice versa. Thus, we use the Hamming weight as cost function $f(.)$ for our \acp{SDM}.
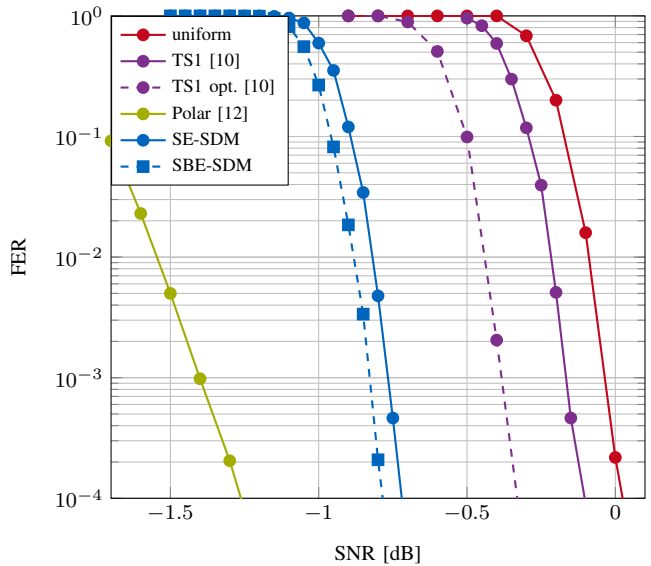
\begin{figure}[t]
    \centering
    \begin{tikzpicture} %
	\begin{axis}[
		height=8cm,
		width=.98\linewidth,
		grid=both,
		xlabel={SNR [dB]},
		ylabel={FER},
		ymode=log,
		xmin=-1.7,
		xmax=0.1,
		ymin=1e-4,
		ymax=1,
		legend cell align={left},
		legend style={at={(0,1)},anchor=north west, font=\scriptsize, cells={align=left}} %
	]
    \footnotesize

    \addplot[thick, plot2, mark=*] table [x=SNR, y=FER] {res/dvbs2_r025_nShpBits=0_uni.txt};
    \label{plot:r025:uniform}
    \addlegendentry{uniform}%

    \addplot[thick, plot4, mark=*] table [x=SNR, y=FER] {res/dvbs2_r050_nShpBits=0_TS1.txt};
    \label{plot:r025:TS1}
    \addlegendentry{TS1\cite{git2019protographbased}}%

    \addplot[thick, plot4, dashed, mark=*, mark options={solid}] table [x=SNR, y=FER] {res/n65k/ldpc_opt_ts1.txt};
    \label{plot:r025:TS1opt}
    \addlegendentry{TS1 opt. \cite{git2019protographbased}}%

    \addplot[thick, plot5, mark=*] table [x=SNR, y=FER] {res/SCL32+CRC32+Lenc32_65536_16384_shp.txt};
    \label{plot:r025:Polar_shp}
    \addlegendentry{Polar \cite{wiegart2019shaped}}

    \addplot[thick, plot1, mark=*] table [x=SNR, y=FER] {res/n65k/dvbs2_r060_nShpBits=3.0k_nJointShp=1_C2_pp_bu_4b.txt};
    \label{plot:r025:SE-SDM}
    \addlegendentry{SE-SDM} %
    
    \addplot[thick, plot1, dashed, mark=square*, mark options={solid}] table [x=SNR, y=FER] {res/n65k/dvbs2_r060_nShpBits=3.0k_nJointShp=5-reshape5.txt};
    \label{plot:r025:SBE-SDM}
    \addlegendentry{SBE-SDM} %

\end{axis}
\end{tikzpicture}%
%
    \caption{\acp{FER} with overall rate $R=1/4$ and $\nfec\approx \num{65000}$.}
    \label{fig:fer_R=0.25}
\end{figure}
We evaluate the end-to-end \acp{FER} and compare with the polar codes in \cite{wiegart2019shaped}, the \ac{TS} scheme in \cite{git2019protographbased}, and \ac{G-SDM} in \cite{wiegart2023probabilistic}.

Fig.~\ref{fig:fer_R=0.67} has an overall rate $R=2/3$ and $\nfec\approx 1000$. The asymptotic shaping gain for this rate is \SI{0.9}{dB}, see \cite[Fig.~1]{wiegart2019shaped}.
We use an unstructured \ac{IRA} code with $\nfec=1056$ and $\Rfec=3/4$. The parity-check matrix was constructed using \ac{PEG} \cite{hu2005regular} and has almost the same degree distributions as the corresponding $\Rfec=3/4$ \ac{LDPC} code of the WiMAX standard \cite{ieee_wimax}, which was also used in \cite{wiegart2023probabilistic}. For \ac{SE-SDM}, we use $\ell=10$ shaping bits. For \ac{SBE-SDM}, we use $\ell=10$ and $\tau=5$.

The \acp{FER} of \ac{SE-SDM} (\ref{plot:r067:SE-SDM}) and \ac{SBE-SDM} (\ref{plot:r067:SBE-SDM}) are almost identical and on par with the \acp{FER} reported in \cite{wiegart2023probabilistic} for both the WiMAX code with $\ell=10$ (\ref{plot:r067:G-SDM_Wimax}) and the 5G LDPC code \cite{5g} with $\ell=32$ (\ref{plot:r067:G-SDM_5G}), for which puncturing improved performance; see \cite{wiegart2023probabilistic}. We gain $\SI{0.2}{dB}$ over \ac{TS} with the $\Rfec=3/4$ WiMAX code (\ref{plot:r067:wimax_TS}), and $\SI{0.9}{dB}$ over uniform signaling with the $\Rfec=2/3$ WiMAX code (\ref{plot:r067:wimax_uniform}), which coincides with the maximum shaping gain. The best reported \acp{FER} are achieved by tailored polar codes from \cite{wiegart2019shaped}, which we depict for $\nfec=1024$, a \SI{16}{bit} outer CRC, and \ac{SCL} encoding and decoding with list size $L_{\text{SCL}}=32$ (\ref{plot:r067:polar}). Note that our code was adapted from the WiMAX code to satisfy the \ac{IRA} structure but was not specifically designed for this channel model. We also plot the finite length \ac{RCU} bound (\ref{plot:r067:RCU}) based on the saddlepoint approximations from \cite{font-segura18_saddlepoint}. The \ac{SBE-SDM} operates within $\approx\SI{0.8}{dB}$ of the \ac{RCU} bound.

Fig.~\ref{fig:fer_R=0.25} depicts \acp{FER} for $R=1/4$ and DVB-S2 \ac{LDPC} codes with $\nfec=\num{64800}$ that possess the \ac{IRA} structure. The \ac{TS} schemes use $\Rfec=1/2$, which is optimal here \cite{git2019protographbased}. However, the best code rate for \ac{SDM} may differ. We conducted a grid search over the code rates $\Rfec\in\lbrace \tfrac{1}{3},\tfrac{2}{5},\tfrac{1}{2},\tfrac{3}{5},\tfrac{2}{3},\tfrac{3}{4}  \rbrace$ defined in the DVB-S2 standard and coarsely optimized $\ell$ for each rate.\footnote{We report the final values of $\ell$, which are typically less than the target values that we optimized over; see the concluding remark in Sec.~\ref{subsec:shaping_bit_selection}.}
The best \ac{SE-SDM} performance was obtained with $\Rfec=0.6$ and $\ell=2627$ shaping bits (\ref{plot:r025:SE-SDM}).
We observe gains of $\SI{0.6}{dB}$ and $\SI{0.75}{dB}$ over \ac{TS} (\ref{plot:r025:TS1}) and uniform signaling (\ref{plot:r025:uniform}), respectively. \ac{SE-SDM} even improves by $\SI{0.4}{dB}$ over \ac{TS} with a tailored code design (\ref{plot:r025:TS1opt}). With \ac{SBE-SDM} with $\tau=5$ and $\ell=2610$ shaping bits (\ref{plot:r025:SBE-SDM}), the total gain compared to uniform signaling increases to $\SI{0.8}{dB}$. 
We do not provide curves for \ac{G-SDM} as the generator matrix was too dense, and encoding with $\ell \approx 2600$ iterations is infeasible in practice.

From \cite[Fig.~1]{wiegart2019shaped}, one can expect shaping gains up to \SI{2}{dB} with optimal \ac{PPS}.
\ac{SBE-SDM} falls short of the optimized polar codes (\ref{plot:r025:Polar_shp}) from \cite{wiegart2019shaped} by $\SI{0.4}{dB}$ at $\text{FER}=10^{-4}$, which realize almost their full potential with shaping gains of $\SI{1.8}{dB}$; see \cite[Fig.~4]{wiegart2019shaped}. However,
\ac{SE-SDM} and \ac{SBE-SDM} show better \ac{FER} slopes, which is why \ac{LDPC} codes are typically preferred over polar codes for these lengths, besides implementation aspects such as throughput and latency. Tailored code design may unlock further shaping gains of up to $\SI{1.2}{dB}$ for \ac{SE-SDM} and \ac{SBE-SDM} in this scenario.

\section{Conclusion}
\label{sec:conclusion}
We proposed \ac{SE-SDM} and a sequential block encoding variant \ac{SBE-SDM} for efficient \ac{PPS} of \ac{IRA} codes. A constructive shaping bit selection strategy for off-the-shelf \ac{IRA} codes was presented. Simulations show that \ac{SE-SDM} and \ac{SBE-SDM} perform on par with the \ac{G-SDM} from \cite{wiegart2023probabilistic} for $\nfec\approx 1000$, and allow for efficient \ac{PPS} of \ac{LDPC} codes with $\nfec=\num{64800}$. We observed gains of $\SI{0.8}{}$--$\SI{0.9}{dB}$ over uniform signaling.

\section*{Acknowledgment}
The authors thank Gerhard Kramer for helpful comments.

\bibliographystyle{IEEEtran}
\bibliography{shaping,ldpc}

\end{document}